# Agentic Literacy Debt: A Structural Problem the AI Literacy Field Has Not Yet Named

**Rohith Nama**

**Abstract**

Autonomous AI agents now plan, decide, and act on behalf of users across healthcare, financial services, and workplace contexts, often without step-by-step human approval. Existing AI literacy frameworks were built for a world in which humans evaluate AI outputs and decide whether to act; they have no vocabulary for the user who has delegated decision-making authority to an agent whose actions may not be observable, reversible, or controllable. This paper names the resulting problem agentic literacy debt: the accumulating societal deficit that grows when agentic AI systems are deployed at scale without corresponding literacy infrastructure. The debt compounds through three reinforcing channels (normalization of opaque delegation, multi-agent ecosystem complexity, and institutional path dependence), and it is incurred by the organizations that deploy agents but paid by the users, patients, and citizens on whose behalf the agents act. Evidence from healthcare, financial fraud, and global equity contexts suggests the gap is already consequential. The problem is structural, not a temporary lag that curriculum reform will close. It demands a reframing of AI literacy as a governance capability, not an evaluative one.

AI literacy frameworks were built for systems that respond. Agentic systems act. That distinction breaks the assumptions every current literacy model depends on. Long and Magerko's (2020) widely adopted definition treats AI literacy as the capacity to evaluate AI outputs, communicate with AI, and use AI as a tool. Every major framework in the decade since has built on that foundation. The foundation assumed a human who queries, reads, and decides. That human is not the human agentic AI creates. When an AI agent executes a sequence of actions across email, calendar, payments, and external services on a user's behalf, the user is no longer an evaluator of anything; the user is a principal who has delegated authority to a system whose actions may not be observable, reversible, or controllable. Existing AI literacy has no vocabulary for this relationship, and the consequences are already visible.

In 2025, security researchers documented real-world prompt injection attacks against users of OpenClaw, a consumer AI agent with more than 150,000 GitHub stars integrating with email, calendars, and terminal execution through messaging apps [1]. A category of attack emerged called "walletdrain": malicious instructions embedded in public social-media posts that caused users' agents, when processing those posts, to execute unauthorized cryptocurrency transactions. The users did nothing wrong by the standards of current AI literacy. They did not generate bad output, fail to evaluate a response, or misuse a tool. They had delegated authority to an agent, and the agent was manipulated through a channel they never saw. No public AI literacy curriculum teaches users to anticipate, monitor, or contest this. The EchoLeak vulnerability in Microsoft 365 Copilot (CVE-2025-32711) demonstrated the same mechanism inside enterprise deployments, causing exfiltration of sensitive organizational data from a single malicious email [2]. These are not edge cases. They are what agentic AI failure looks like.

Existing AI literacy rests on three assumptions that agentic AI invalidates. It assumes evaluation is possible: the user sees an output and judges it. Agentic systems produce action chains most of which the user never observes. It assumes reversibility: a bad decision based on AI advice can be reconsidered before consequence. Agentic systems can transfer money, send messages, and delete records in milliseconds, none of it recoverable by user competence after the fact. And it assumes control: the user remains the agent of action, the AI merely an informant. Agentic systems invert this, acting as the agent of action while the user becomes a principal with authority but without visibility. Evaluation breaks down, oversight becomes necessary in its place, and responsibility shifts away from the user who pressed a button and toward a distributed set of parties, any of which may or may not have controlled the specific decision that caused harm. The failure is not incremental. Three assumptions break at once, and no existing literacy framework addresses any of them.

I call this accumulating gap agentic literacy debt. Borrowed from software engineering's notion of technical debt, where expedient short-term decisions create compounding future costs, agentic literacy debt accumulates when agentic AI systems are deployed at scale without the literacy infrastructure users need to govern them. Every deployment of an autonomous agent for healthcare scheduling, financial planning, enterprise productivity, or personal assistance, without frameworks enabling users to understand, supervise, and contest its actions, adds to the debt. Like technical debt, it compounds, but the compounding mechanism is specific. It operates through at least three reinforcing channels. First, normalization: each opaque delegation habituates users to granting permissions without scrutiny, lowering the threshold for the next. Permission grants in production agentic systems are typically inherited across sessions and rarely revoked, producing a ratchet effect where each interaction quietly expands the agent's access surface. Second, ecosystem complexity: each new agent interacts with previously deployed agents and services, producing multi-agent chains that are harder to oversee than any individual system. Third, institutional path dependence: organizations that skip literacy infrastructure for one deployment build no capacity to provide it for the next, and the cost of retrofitting grows with each iteration. The result is not merely an additive gap but a widening structural deficit.

The metaphor of compounding debt applied to structural inequality has precedent. Ladson-Billings [3] reframed educational disparities in the United States by arguing that annual "achievement gaps" are merely snapshots of a cumulative "education debt" with historical, economic, sociopolitical, and moral dimensions. Her insight, that focusing on annual gaps obscures the structural, compounding nature of the problem, applies directly to AI literacy. Petrozzino [4], writing in AI and Ethics, introduced the concept of "ethical debt" in AI: the cost incurred when organizations deploy AI systems without proactively identifying ethical concerns. She made the critical observation that those who incur the debt (developers and deploying organizations) differ from those who pay for it (marginalized communities). Agentic literacy debt extends both. Where Ladson-Billings addresses cumulative failures of educational systems and Petrozzino addresses ethical shortcuts during AI design, agentic literacy debt describes the accumulating societal deficit in the knowledge, skills, and institutional frameworks required to govern AI systems that act autonomously on citizens' behalf. Unlike ethical debt, which is incurred at the point of design, it is incurred at the point of deployment and compounds with every user interaction that occurs without adequate literacy infrastructure.

Critically, the debt is not borne by the organizations that incur it. It is paid by the users, patients, consumers, and citizens the agents act upon. This asymmetry between who creates the debt and who pays for it is what makes agentic literacy debt an AI ethics problem rather than merely an educational one.

## The Costs of This Gap Are Already Visible

Direct empirical evidence of agentic-specific harms is still emerging; the sections that follow draw on evidence from adjacent domains to illustrate the trajectory rather than to document the endpoint, and should be read as illustrative rather than as systematic evidence.

**Healthcare.** Agentic AI systems for patient triage, scheduling, and care navigation are expanding rapidly, with the global market reaching $538 million in 2024 and projected to grow at 45.56% annually [5]. Trust dynamics documented in clinical AI research, though drawn from predictive and diagnostic systems rather than agentic ones, are directly relevant: a systematic review found AI explainability increased clinician trust in half of studies and decreased it in others depending on explanation quality [6], and general users systematically overtrust AI-generated medical advice despite low accuracy [7]. When AI systems move from advising clinicians to acting on patients' behalf, these trust calibration failures will carry greater stakes and offer fewer opportunities for human correction.

**Fraud and security.** Reported fraud losses reached $12.5 billion (FTC) and $16.6 billion (FBI IC3) in 2024 [8], with Deloitte projecting GenAI-enabled losses in U.S. banking could reach $22 to $40 billion by 2027 [9]. These figures cover all fraud, not agentic AI specifically; what matters is the mechanism shift. Attacks increasingly target not human users but the AI agents acting on their behalf. Indirect prompt injection was rated by OWASP as the number-one risk in production LLM applications [10], and a Gartner survey of 302 cybersecurity leaders found 32% of organizations had already experienced a prompt-based attack [11]. No public AI literacy curriculum addresses this vector. The attack surface bypasses the user entirely, targeting the agent's processing of external content. Users cannot apply literacy skills they were never taught, against threats that operate in a channel they cannot observe.

**Global equity.** In the Asia-Pacific region, only 49% of rural residents use the internet compared to 83% of urban residents, a gap that has widened since 2021 [12]. The GSMA identifies digital skills as the single greatest barrier to mobile internet adoption in Asian countries surveyed [13]. Yet agentic AI is actively deployed in these settings, often precisely because its automation advantages are most compelling in resource-constrained environments. A 2024 systematic review found that no AI literacy measurement scale has been tested for cross-cultural validity [14]. The populations most exposed to the risks of the literacy gap are also those least served by the research designed to close it.

## Why the Gap Is Structural, Not Temporary

The deployment-literacy gap cannot be closed by conventional curriculum reform alone, and the reason is structural. National curriculum updates typically unfold over five to seven years; university program redesign often takes several years from proposal to implementation. Agentic AI capabilities, by contrast, evolve on product cycles measured in months. The result is a structural mismatch between the speed of technological deployment and the pace at which institutional learning systems adapt. This is not a temporary lag. It is a permanent condition unless the delivery mechanism for literacy is itself redesigned around the same technological substrate creating the gap.

Part of the challenge is conceptual. Existing AI literacy frameworks largely focus on helping individuals interpret and evaluate AI-generated outputs. Agentic AI changes the locus of responsibility: instead of assessing information produced by a system, users must supervise actions performed by systems acting on their behalf. They need to set boundaries, monitor decisions, and intervene when processes deviate from intended goals. This absence is consistent across the field's major instruments: UNESCO's 2024 AI Competency Framework for Students, the Meta AI Literacy Scale (MAILS), and the recently proposed AI Literacy Heptagon each define competency domains for evaluating and using AI; none include delegation, oversight, or accountability attribution for autonomous agent action. Recent work on AI agent autonomy formalizes this shift: Feng et al. [15] propose five levels of agent autonomy defined by user roles, and Kasirzadeh and Gabriel [16] construct four-dimensional agentic profiles for proportional governance. Both describe governance challenges but not the literacy infrastructure required for populations to exercise the governance roles they prescribe. Part of the challenge is also architectural. Every production agentic system generates detailed action logs, but these are designed for developer debugging, not user comprehension. The authorization flow through which a user grants an agent access to email, calendar, or financial accounts is typically a single “Allow” button with no scope granularity, no explanation of what the agent can do with that access, and no visible mechanism for revocation. The literacy gap is not only conceptual. It is designed in [17].

## The Ethical Obligation to Close the Debt

The EU AI Act's Article 4, applicable since February 2025, creates the world's first binding AI literacy obligation, requiring providers and deployers to ensure a "sufficient level of AI literacy" among persons affected by AI systems [18]. Yet the Article's implementing guidance does not specify what literacy means in agentic contexts, where the human is no longer evaluating outputs but governing autonomous action. Floridi and Cowls' principles of autonomy and explicability [19] point to the same obligation from another direction: meaningful consent, accountability, and redress depend on the public's capacity to understand and govern systems acting on their behalf. Floridi has argued that AI represents an unprecedented divorce between agency and intelligence, for the first time allowing entities to act consequentially in the world without understanding, consciousness, or intentionality [20]. When such entities act autonomously on behalf of individuals, the absence of public capacity to understand and govern them is not a minor gap; it is the condition that agentic literacy debt describes. Treating it as an educational matter to be addressed in time misreads the structure of the problem.

At minimum, the transition from generative to agentic AI demands new principal-side competencies in:

- delegation (understanding what authority one is granting)
- oversight (monitoring and constraining agent actions)
- accountability attribution (understanding who is responsible when harm occurs)
- attack surface awareness (recognizing that agents can be manipulated through the data they process)
- agent-specific informed consent (knowing when an agent rather than a human is acting)
- calibrated trust (recognizing when both undertrust and overtrust represent failures of informed engagement)

These are not refinements of existing AI literacy. They are structurally new competencies for a structurally new paradigm, and their full specification, including proficiency levels and design imperatives, is the subject of a companion paper [21].

## A Call to the Field

Agentic AI is not merely changing what machines can generate; it is changing what machines can decide and execute. The concept is real, it is compounding, and it is being paid by people who had no say in incurring it.

The concept does not imply that education alone can resolve the risks of agentic AI, but rather that literacy must evolve alongside governance, system design, and regulatory oversight. I call on AI literacy researchers to prioritize competency development for agentic contexts; on AI ethics scholars to examine the informed consent failures implicit in current deployment practices; and on the engineers building these systems to recognize that design defaults are literacy interventions whether or not they are intended as such. The default authorization flow for every major agentic platform today presents users with a single undifferentiated permission grant. The default action log is invisible to the user. The default for irreversible actions is to execute first and report after. Each of these defaults is a decision about whether users will develop the capacity to govern the agents acting on their behalf, or whether they will remain permanently dependent on systems they cannot see, constrain, or contest. The technology creating the debt is capable of helping close it, whether through transparency-by-design that embeds literacy into agent interactions, AI tutoring that simulates agentic scenarios at scale, or contextual micro-learning at the point of risk. But these are not features that emerge organically from product roadmaps optimized for task completion. They require treating user literacy as a first-class design objective.

Without this reframing, agentic AI will be deployed at scale into environments where users are structurally unprepared to govern it, and the debt will be paid by the people least equipped to recognize what they are being charged for.

## Declarations

**Funding:** No funding was received for conducting this study.

**Competing interests:** The author has no financial or non-financial competing interests to declare that are relevant to the content of this article. This work was conducted in the author's personal capacity and does not represent the position of Amazon, or any other organization with which the author is affiliated.

**Data availability:** This is a conceptual paper. No primary data was collected or analyzed.

**Use of AI tools:** AI tools were used for language refinement and structural editing. All conceptual development, analysis, and conclusions are the author's own.